\documentclass[twocolumn,aps,prc,superscriptaddress,showpacs,floatfix, nofootinbib,preprintnumbers]{revtex4}

\usepackage{array}
\usepackage{epsfig}
\usepackage{amsmath}
\usepackage{multirow}
\usepackage{graphicx}
\usepackage{amsmath}
\usepackage{feynmf}
\usepackage[normalem]{ulem}  
\usepackage[dvips]{color} 

\renewcommand\sout{\bgroup \color{red} \ULdepth=-.5ex \ULset}

\newcommand{\Ex}[2]{\ifmmode{#1\times10^{#2}}\else{$#1\times10^{#2}$}\fi}

\begin{document}
\preprint{YITP-15-82}
\title{Temperature dependence of dimension 6 gluon operators and their effects on Charmonium}

\author{HyungJoo Kim}\affiliation{Department of Physics and Institute of Physics and Applied Physics, Yonsei
University, Seoul 120-749, Korea}
\author{Kenji Morita}\affiliation{Yukawa Institute for Theoretical
Physics, Kyoto University, Kyoto 606-8502, Japan}
\author{Su Houng Lee}\affiliation{Department of Physics and Institute of Physics and Applied Physics, Yonsei
University, Seoul 120-749, Korea}
\date{\today}
\begin{abstract}
 Starting from an earlier representation of the independent dimension 6
 gluon operators in terms of color electric and magnetic fields, we
 estimate their changes near the critical temperature $T_c$ using 
 the temperature dependence of the dimension 4 electric and magnetic
 condensates extracted from pure gauge
 theory on the lattice. We then improve the
 previous QCD sum rules for the $J/\psi$ mass near $T_c$ based on
 dimension 4 operators, by including the contribution of the dimension 6
 operators to the OPE. We find an enhanced stability in the sum rule and
 confirm that the $J/\psi$ will undergo an abrupt change in the property
 across $T_c$.
\end{abstract}


\maketitle

\section{Introduction}

The temperature dependence of the gluon condensate, or the trace anomaly
in the pure gauge theory, offers a useful picture on the nature of the
QCD phase transition~\cite{Boyd:1996bx}.
The scalar gluon condensate together with the twist-2 gluon operator are
the two independent gluon operators at dimension 4.
These operators can be re-expressed in terms of the electric and the
magnetic condensate.  The temperature dependence of these operators can
be calculated directly from lattice calculations of the space-time and
space-space elementary plaquette~\cite{Lee:1989qj,Boyd:1996bx}, or from
combining the calculation of the energy density and pressure~\cite{Lee:2008xp}.
The calculations show that while there is a rapid change of the electric
condensate across the phase transition temperature, the magnetic
condensate changes very little~\cite{Lee:2008xp}, somehow reflecting a
possible link to the observed sudden change in the space-time Wilson
loop and the persistent area-law behavior of the space-space Wilson loop
across the phase transition~\cite{Manousakis:1986jh}.

Using the temperature dependence of the dimension 4 condensates as the
input in the QCD sum rule approach for the heavy quark system,  $J/\psi$
and $\eta_c$ have been found to undergo a rapid property change across the
phase transition~\cite{Morita:2007pt,Morita:2007hv,Lee:2008xp,Morita:2009qk} and to their
dissociation~\cite{Gubler:2011ua,Dominguez} slightly above  the critical temperature.

In a recent work, two of us (H.K. and S.H.L.) have performed the
renormalization of all the independent dimension 6 spin-2 gluon
operators and found the  scale invariant combination in the pure gauge
theory~\cite{Kim:2015ywa}.  Together with the dimension 6 scalar
condensate, whose renormalization has been worked out
before~\cite{Narison:1983kn,Tarrach}, the renormalization of the two
dimension 6 gluon operators are now understood.

Unfortunately, a direct lattice calculation is presently not feasible to
extract their temperature dependence since the higher dimensional operators
will come with large uncertainties and the mixing with lower dimensional
operators becomes problematic.
At the same time, in the pure gauge theory, it is interesting to note
that  while the dimension 6 scalar operator is composed of a higher product of gluon
field strength tensor, the  spin-2 part is the second moment of the dimension 4 scalar  gluon condensate, thus suggesting a strong correlation with the dimension 4 operator.   The connections  become more apparent when we express the
dimension 6 operators in terms of color electric ($E$) and magnetic ($B$)
fields.

In this paper, using these expressions and the temperature dependence of
dimension 4 electric and magnetic condensates extracted from lattice
gauge theory, we estimate the changes of the dimension 6 operators near the critical temperature
$T_c$.  Furthermore, we then improve the previous QCD sum rules for
$J/\psi$ mass near $T_c$, which was based on dimension 4 operators, by
including the contribution of the dimension 6 operators to the OPE,
whose nuclear medium effect was previously studied in
Ref.~\cite{SuHoungLee}.
We find an enhanced stability in the sum rule and confirm that the
$J/\psi$ will undergo an abrupt change in the property across
$T_c$~\cite{Morita:2007pt,Morita:2007hv}.

In section II, we will show $E$ and $B$ fields representations of dimension
6 gluon operators. In section III, we will show the temperature dependence
of the condensates.  In section IV, we will apply our condensates to the
sum rules. Section V is devoted to a summary.

\section{Field Representation}
At dimension 4, there are two independent gluon operators that can be
constructed  from $G^a_{\mu\alpha}G^a_{\nu\alpha}$. These are the scalar
and the twist-2 gluon operators.
\begin{flalign}
&g_{\mu \nu} [G^a_{\mu \alpha}G^a_{\nu \alpha}]=G^a_{\mu \nu}G^a_{\mu\nu} \nonumber \\
&G^a_{\mu \alpha}G^a_{\nu\alpha}|_{ST}=(u_{\mu}u_{\nu}-g_{\mu\nu}/4)G_2, \label{dim4-1}
\end{flalign}
where the subscript $ST$ represents symmetric and traceless
indices. These operators can be represented by $E$ and $B$ fields as
follows.
\begin{flalign}
&G^a_{\mu \nu}G^a_{\mu\nu}=2(B^2-E^2) \nonumber \\
&G_2=-\frac{2}{3}(E^2+B^2),
\end{flalign}
where the trace is taken for the color indices of $E$ and $B$ fields and
the medium four vector is taken to be $u^\mu=(1,0,0,0)$ in this work.

For dimension 6 operators, in addition to the twist-2 gluon
operator, there are two more independent  dimension 6 gluon operators
that remain after using the equations of motion in the pure gauge
theory. Introducing the short hand notation
$G^3_{\mu \nu}\equiv f^{abc}G^a_{\mu \alpha}G^b_{\alpha
\beta}G^c_{\beta\nu} $,
one finds these operators in a similar form to the dimension 4 case,
\begin{flalign}
&g_{\mu \nu} [G^3_{\mu \nu}]=f^{abc} G^a_{\mu \alpha} G^b_{\alpha \beta} G^c_{\beta\mu}\nonumber \\
&G^3_{\mu \nu}|_{ST}=(u_{\mu}u_{\nu}-g_{\mu\nu}/4)G_3.    \label{dim6-0}
\end{flalign}
These operators can also be represented by the $E$ and $B$ fields. Using  parity and rotational symmetry, we find that  $E^a E^b E^c$
and $B^a B^b E^c$ type of operators vanish such that the remaining forms
are given as follows:
\begin{flalign}
g_{\mu \nu} [G^3_{\mu\nu}]&=f^{abc}[3B^a\cdot (E^b\times E^c)-B^a\cdot(B^b\times B^c)] \nonumber \\
G_3&=\frac{f^{abc}}{3} B^a \cdot(B^b \times B^c +E^b \times E^c).  \label{dim6}
\end{flalign}
In the following we will abbreviate  the triple scalar product of fields $f^{abc} A^a \cdot (B^b \times C^c)$ as $ABC$ for simplicity.  Using this notation,
the first and second line of Eq.~(\ref{dim6}) will be represented as $[3BEE-BBB]$ and $[BBB+BEE]/3$ respectively.


\section{Temperature dependence of condensates}

We start from the temperature dependence of  $\langle
\frac{\alpha_s}{\pi}E^2 \rangle_T$ and $\langle \frac{\alpha_s}{\pi}B^2
\rangle_T$ extracted from lattice calculations~\cite{Lee:2008xp} and
discuss how the temperature dependence of higher dimensional operators
can be estimated.

The temperature dependent dimension 4 condensates can be expressed as follows.
\begin{eqnarray}
& \displaystyle \left\langle \frac{\alpha_s}{\pi}G^a_{\mu \alpha}G^a_{\mu \alpha} \right\rangle_T&=2\left[\left\langle \frac{\alpha_s}{\pi}B^2 \right\rangle_T -\left\langle \frac{\alpha_s}{\pi}E^2 \right\rangle_T\right] \\
& \displaystyle \left\langle \frac{\alpha_s}{\pi}G_2 \right\rangle_T&=-\frac{2}{3}\left[\left\langle \frac{\alpha_s}{\pi}B^2 \right\rangle_T +\left\langle \frac{\alpha_s}{\pi}E^2 \right\rangle_T\right].
\end{eqnarray}

For the dimension 6 condensates, we first note that all dimension 6
operators with either spin 0 or 2 can be represented as linear
combinations of the operators in Eq.~\eqref{dim6-0} in pure gauge
theory~\cite{SuHoungLee}, so that we can represent the operators with two covariant derivatives as
\begin{align}
\left\langle \frac{\alpha_s}{\pi}G^a_{\mu \nu}G^a_{\mu\nu;\kappa\kappa}
 \right\rangle_T & =  2 \left\langle \frac{g\alpha_s}{\pi}G^3_{\mu \mu} \right\rangle_T \hfill
 \nonumber \\
 =
\frac{4}{\pi^{1/2}}&[3\langle \alpha^{3/2}_sBEE \rangle_T -\langle
 \alpha^{3/2}_sBBB \rangle_T], \label{dim6-1}
\end{align}
\begin{align}
 X_T& = 2 \left\langle \frac{g \alpha_s}{\pi} G_3 \right\rangle_T \nonumber \\
 & =  \frac{4}{3\pi^{1/2}}[\langle \alpha^{3/2}_sBEE \rangle_T +\langle \alpha^{3/2}_sBBB \rangle_T], \label{dim6-2}
\end{align}
where we have defined
\begin{eqnarray}
\left\langle \frac{\alpha_s}{\pi}G^a_{\kappa \lambda}G^a_{\kappa\lambda;\mu\nu} |_{ST}\right\rangle_{T}  =
\left(u_\mu u_\nu - \frac{1}{4}g_{\mu \nu}\right)X_T.
\end{eqnarray}

\begin{figure}[!t]
  \centering
  \includegraphics[width=0.41\textwidth]{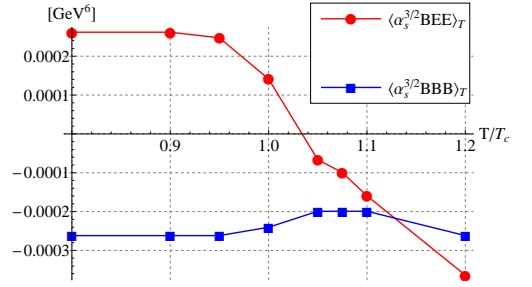}\\
  \caption{Temperature dependence of BBB and BEE.}
  \label{fig1}
\end{figure}

To estimate the temperature
dependencies of $BEE$ and $BBB$ from temperature dependencies of the
lower dimensional field operators $E^2$ and $B^2$, we assume that fields
are isotropic and that the angular correlations can be
neglected. Namely, $|E_i^a| = E$ or 0 and $|B_i^a| = B$ or 0 for
$a=1,\cdots,8$. Such an assumption is actually satisfied in an
instanton configuration where $E_i^a, B_i^a \propto
\delta_{ai}$ \cite{Belavin:1975fg, Schafer:1996wv} and is used
successfully in the vacuum \cite{Nikolaev:1982rq}. \footnote{It is not
trivial that such a configuration still holds at high temperature. However, since it gives the maximum of the triple scalar product, our estimate can  be
regarded as conservative upper bound.}

Hence, we approximate
\begin{flalign}
&\langle \alpha^{3/2}_sBEE \rangle_T=\langle \alpha^{3/2}_sBEE \rangle_0 \frac{{\langle \frac{\alpha_s}{\pi}B^2 \rangle_T}^{1/2}\langle \frac{\alpha_s}{\pi}E^2 \rangle_T}{{\langle \frac{\alpha_s}{\pi}B^2 \rangle_0}^{1/2}\langle \frac{\alpha_s}{\pi}E^2 \rangle_0}\\
&\langle \alpha^{3/2}_sBBB \rangle_T=\langle \alpha^{3/2}_sBBB \rangle_0 \frac{{\langle \frac{\alpha_s}{\pi}B^2 \rangle_T}^{3/2}}{{\langle \frac{\alpha_s}{\pi}B^2 \rangle_0}^{3/2}}.
\end{flalign}
Furthermore, using the fact that $fG^3$=$4BEE$=$-4BBB$ in the Euclidean spacetime at zero temperature and approximating
 $\langle g^3fG^3\rangle_0$=$(0.6 \text{GeV})^6$~\cite{Nikolaev:1982rq}, we can determine the vacuum
values at zero temperature.
\begin{eqnarray}
\langle \alpha^{3/2}_sBEE \rangle_0 =-\langle \alpha^{3/2}_sBBB
 \rangle_0=\frac{(0.6 \text{GeV})^6}{4(4\pi)^{3/2}}.
\end{eqnarray}
The temperature dependences of Eqs.~(10) and (11) are then given in Fig.~\ref{fig1}.
Furthermore, the temperature dependence of the operators in
Eq.~(\ref{dim6-1}) and Eq.~(\ref{dim6-2}) are accordingly  obtained.

\section{Application to sum rules for $J/\Psi$}

\begin{figure*}[!t]
  \centering
  \includegraphics[width=0.49\textwidth]{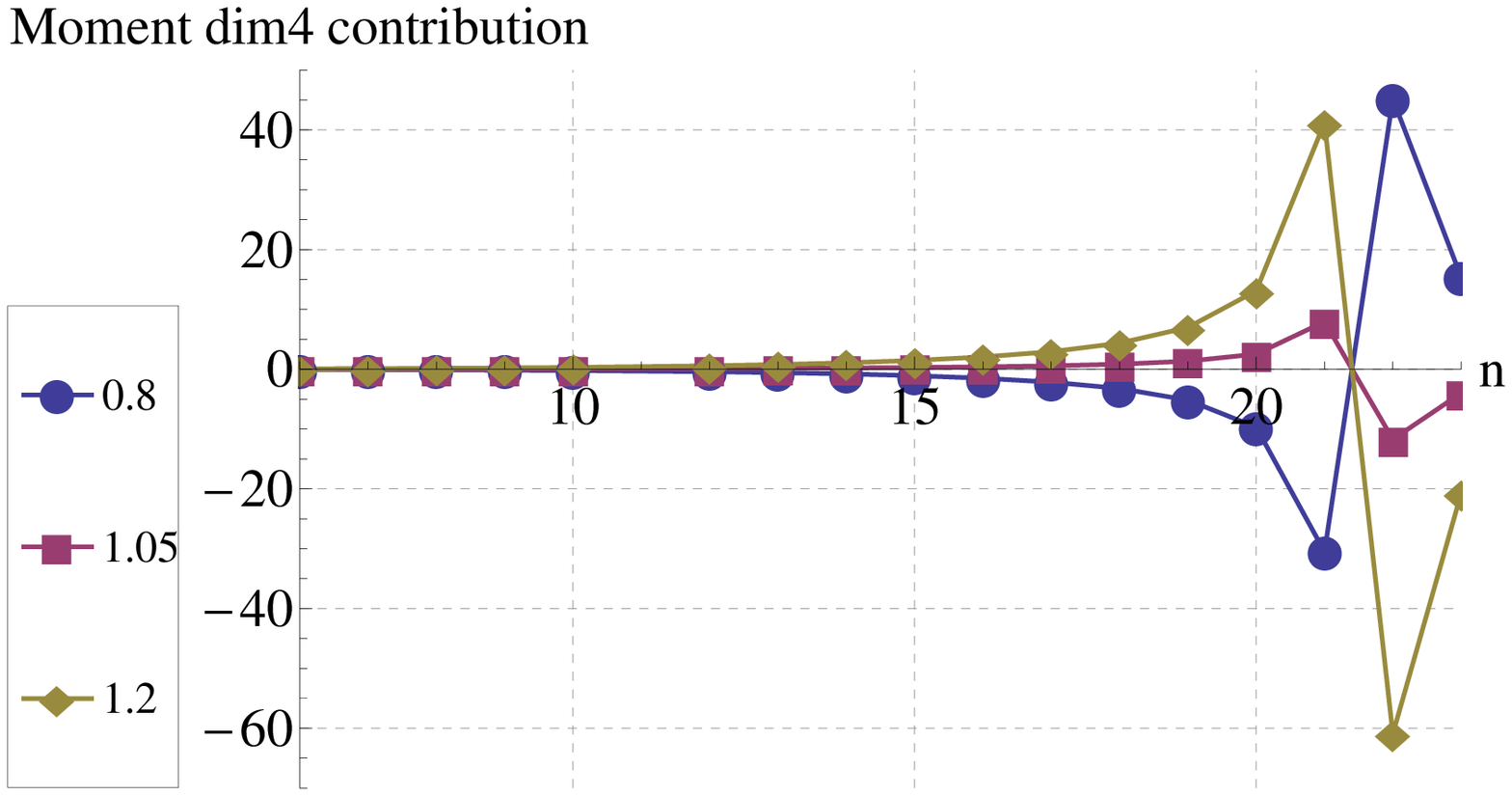}
  \includegraphics[width=0.49\textwidth]{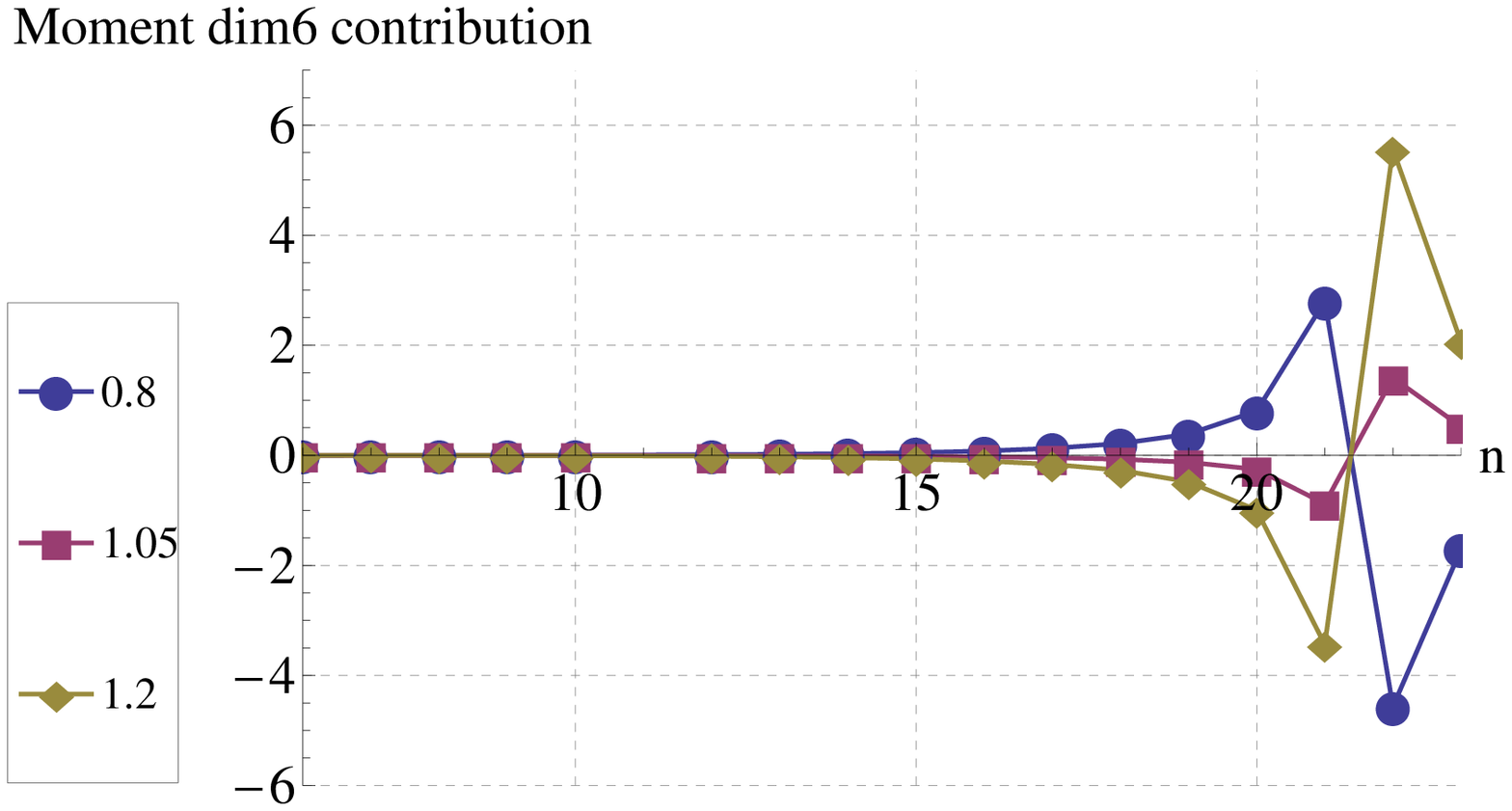}\\
  \caption{The left and right shows the contributions to $M_n(\xi)$ from dimension 4 and dimension 6 condensates, respectively,  divided by perturbative contribution. The circle, square and diamond represent $T=0.8,1.05$ and $1.2T_c$ respectively.}
  \label{fig2}
  \includegraphics[width=0.49\textwidth]{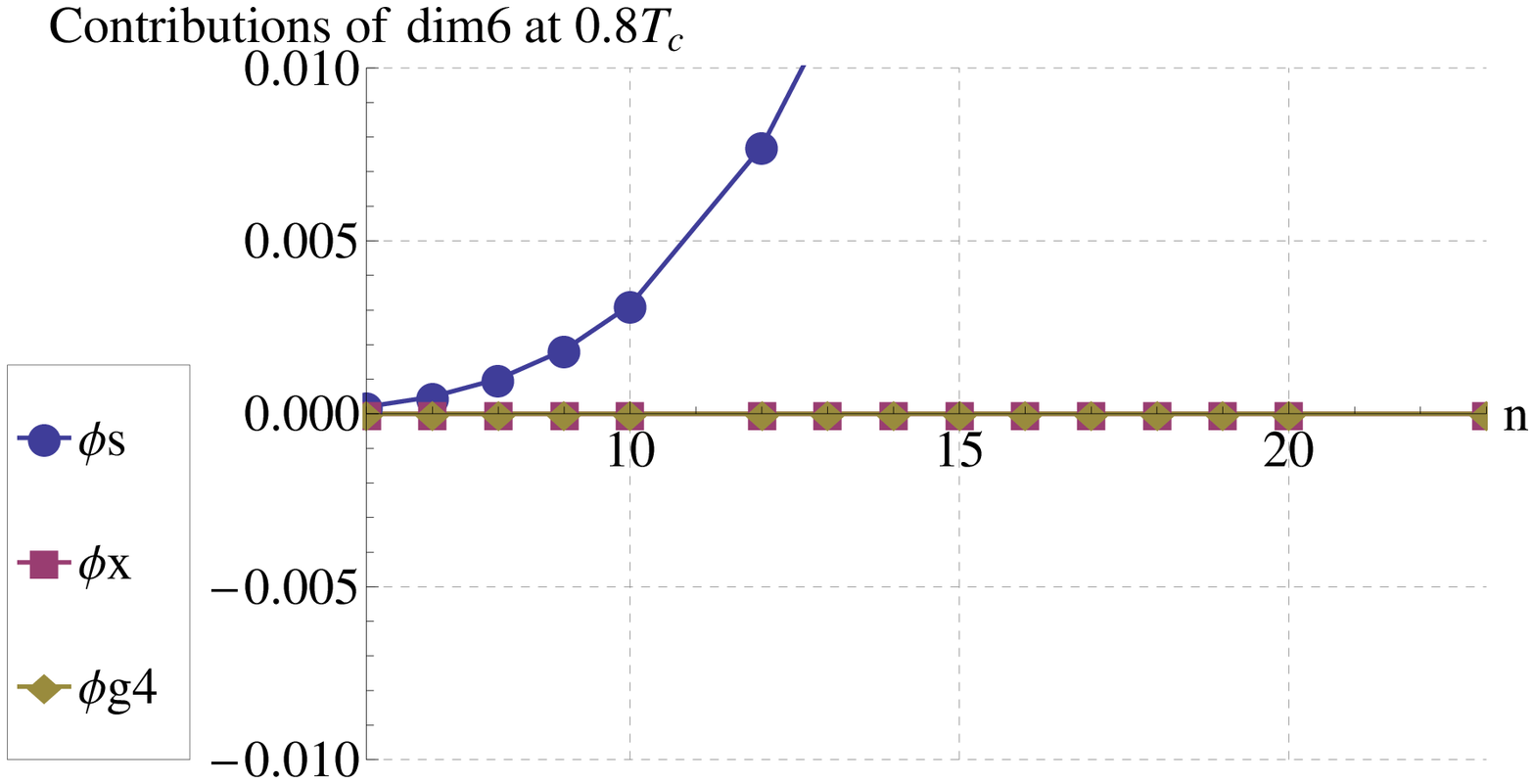}
  \includegraphics[width=0.49\textwidth]{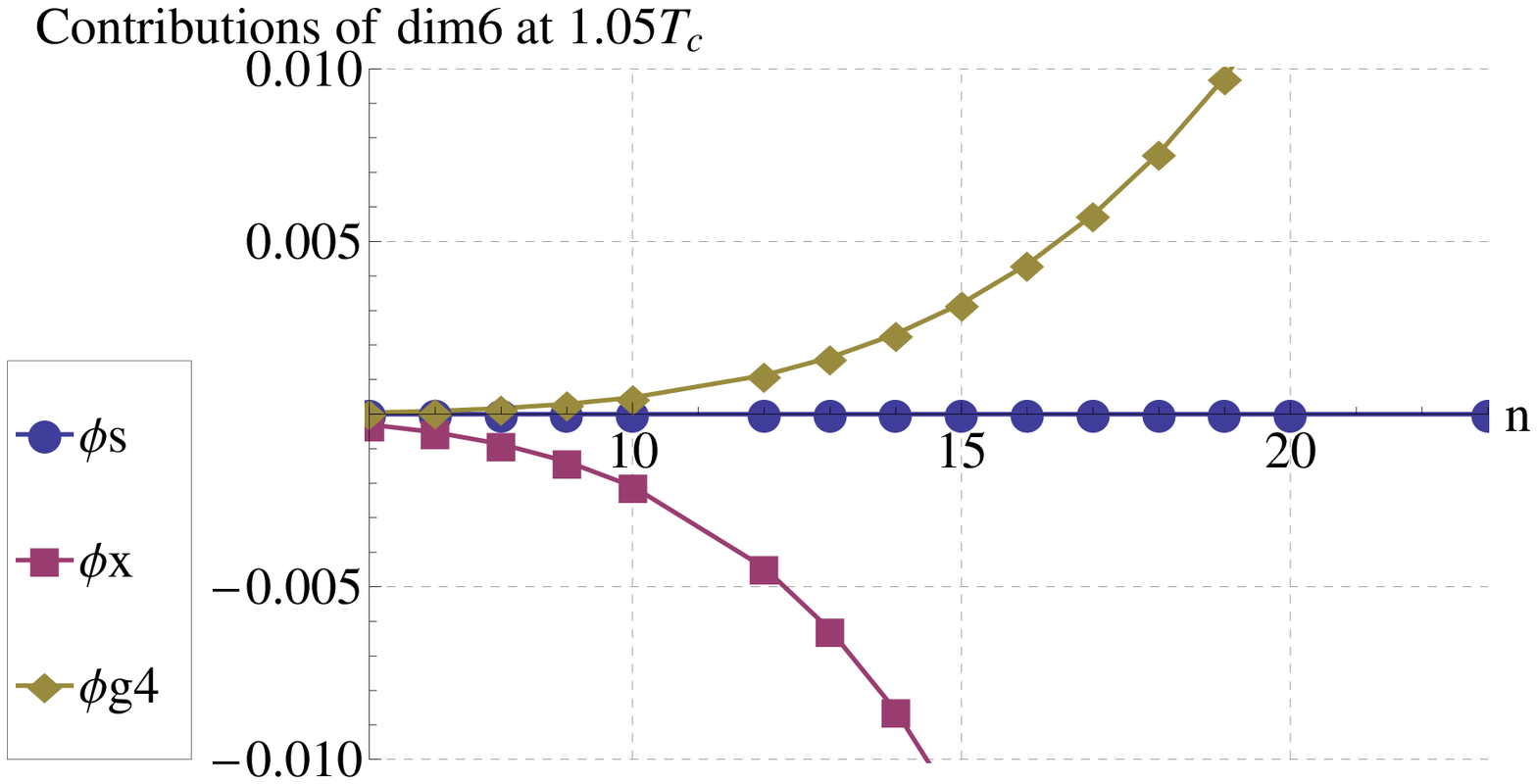}\\
  \caption{Contributions to $M_n(\xi)/A^V_n(\xi)$ of dimension 6 condensates at $0.8T_c$(left figure) and at $1.05T_c$(right figure).}
  \label{fig3}
  \includegraphics[width=0.48\textwidth]{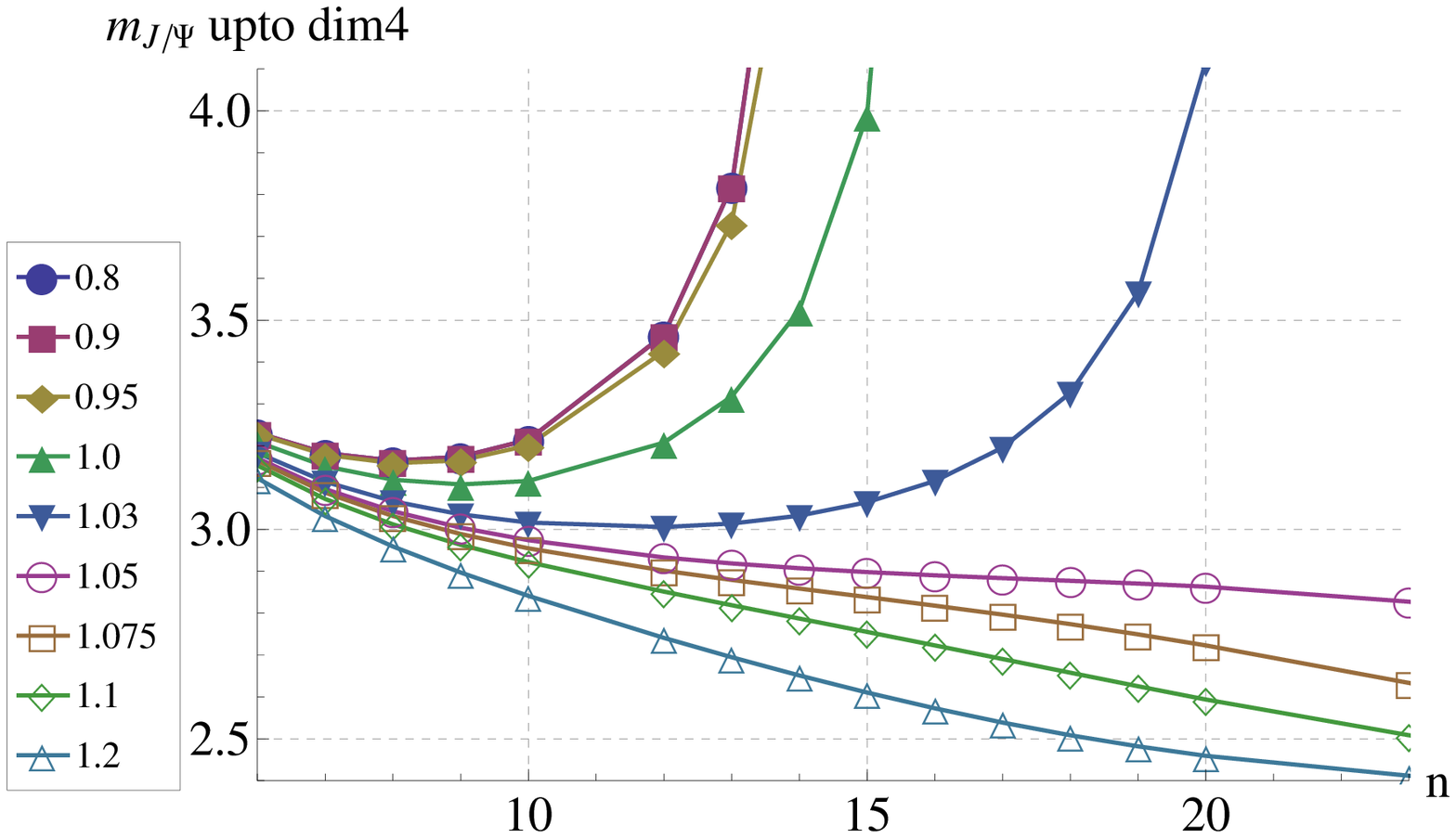}
  \includegraphics[width=0.48\textwidth]{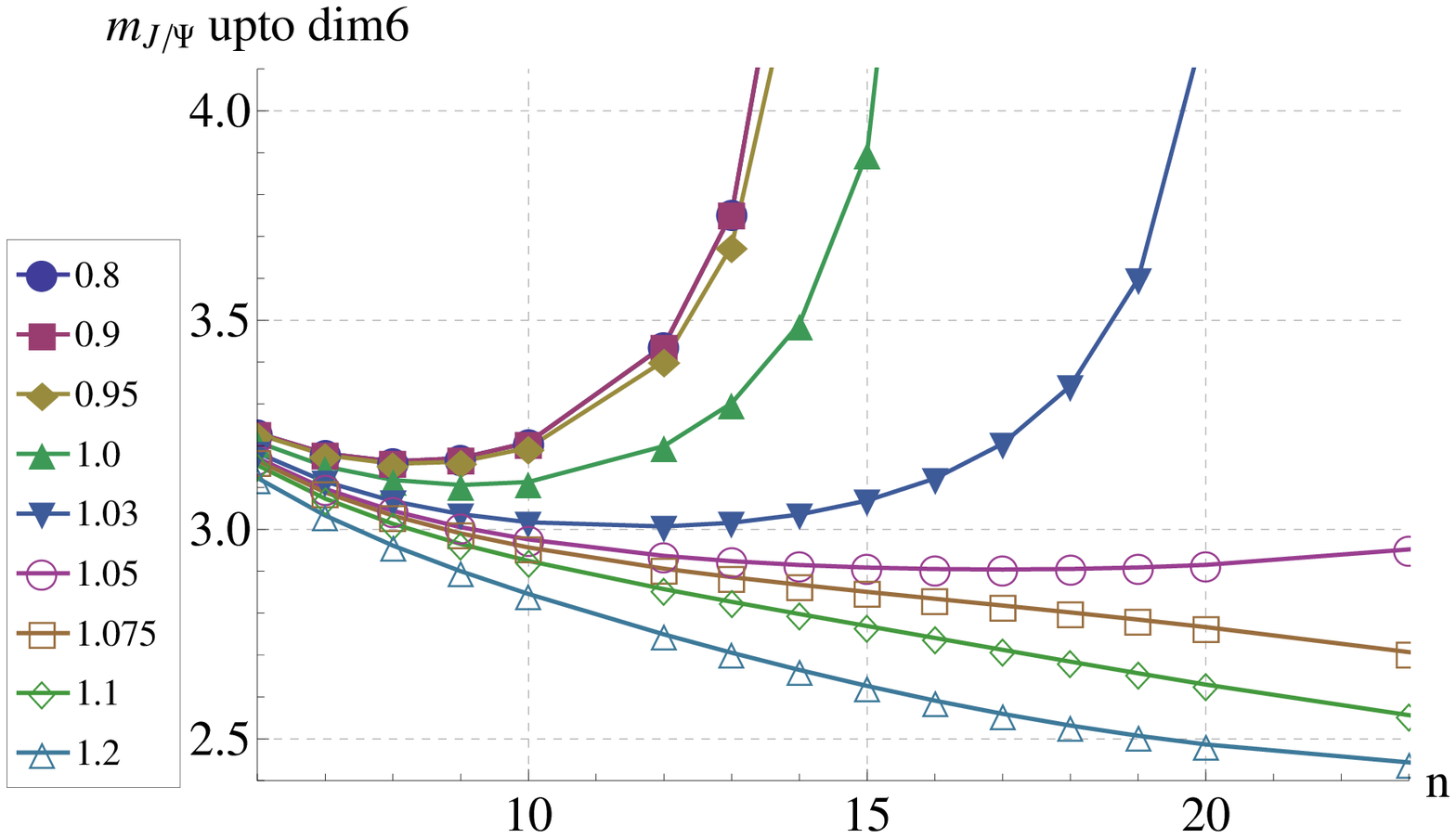}\\
  \caption{Temperature dependence of $m_{J/\psi}$. Left:  considering up
 to dim4 condensates. Right: up to dim 6 condensates.}
  \label{fig4}
\end{figure*}

Using the temperature dependence of the dimension 4 gluon operators, two
of us have calculated the mass of $J/\psi$ using QCD moment sum rules
near the critical temperature~\cite{Morita:2007pt,Morita:2007hv}.
There, it was found that the properties of the $J/\psi$ underwent a
sudden change at $T_c$.
However, the sum rule was also found to become unstable above $1.05 T_c$. This instability was later found to be linked to an onset of the broadening~\cite{Morita:2009qk} and to a precursor effect of the melting
of $J/\psi$ at slightly higher temperature as being found by application of the Maximum Entropy method to the sum rule~\cite{Gubler:2011ua}.  To confirm the QCD
sum rule analysis with better stability, we generalize the sum rule
analysis to include the contributions from dimension 6 operators.

For that purpose, we have to further include the temperature dependence of
dimension 6 twist-2 operator, which we have not discussed so far. This
operator can not be expressed in terms of $E$ and $B$ fields directly and is
parametrized as follows:
\begin{align}
& \left\langle \frac{\alpha_s}{\pi}G^a_{\mu
 \kappa} G^a_{\nu\kappa;\alpha\beta} |_{ST} \right\rangle  \nonumber\\
& \quad =G_4 [u_\mu u_\nu u_\alpha u_\beta+\frac{1}{48}(g_{\mu\nu}
 g_{\alpha\beta}+g_{\mu\alpha}g_{\nu\beta}+g_{\mu\beta} g_{\nu\alpha})
 \nonumber\\
& \quad -\frac{1}{8}(u_\mu u_\nu g_{\alpha\beta}+u_\mu u_\alpha
 g_{\nu\beta}+u_\mu u_\beta g_{\alpha\nu}+u_\nu u_\alpha g_{\mu\beta}
 \nonumber \\
& \quad + u_\nu u_\beta g_{\mu\alpha}+ u_\alpha u_\beta g_{\mu\nu})].
\end{align}
It should be noted that $G_4$ is the higher moment of the dimension 4
twist-2 $G_2$ operator in Eq.~(\ref{dim4-1}).
To estimate the value of
$G_4$ we use the estimate near $T_c$ based on a quasiparticle picture
given in Ref.\cite{Morita:2010pd}. There, the ratio between $G_4$ and
$G_2$ is given by the square of the mass of the quasiparticle times $A_4/A_2$,
where $A_i$ is the $i$-th moment of the structure function of the quasiparticle.  Hence we
will assume
\begin{align}
{G_4}/{G_2}\sim -m_G^2 A_4/A_2. \label{g4-1}
\end{align}
For the numerical value, we take $m_G=0.6$ GeV for the thermal gluon
mass near $T_c$ and take $A_4=0.02$ and $A_2=0.9$ for a typical value
for any hadron as discussed in given in Ref.~\cite{SuHoungLee}; such approximation  should be valid below $T_c$.  To extend the formula to temperature above $T_c$, we make use of the thermal fluctuation of the thermal gluon momentum assuming a temperature dependent thermal mass as extracted from lattice gauge theory~\cite{Levai:1997yx}.  Noting that $G_4$ involves two additional covariant derivatives in comparison to $G_2$, we will use the following approximation.
\begin{align}
&{G_4}/{G_2} \sim -\left(m_G^2\frac{A_4}{A_2}\right)\frac{{\left\langle p^2\right\rangle}_T}{{\left\langle p^2\right\rangle}_{T_c}},
\end{align}
where we take
\begin{align}
 \frac{\langle p^2 \rangle_T}{\langle p^2 \rangle_{T_c}} =\left.
 \frac{\int_{0}^{\infty} dp p^4 n_B(m_\text{eff}(T),
 T)}{\int_{0}^{\infty} dp p^4 n_B(m_\text{eff}(T_c), T_c)}\right\vert_{T_c=260\text{MeV}}
\end{align}
with $n_B(m,T)$ being the Bose distribution function
$(e^{\sqrt{p^2+m^2}/T}-1)^{-1}$ for gluons
and
\begin{align}
m_\text{eff}(T)=0.6+0.062\left(\frac{T}{T_c}-1.5\right)^2 \, \text{GeV},
\end{align}
as parametrized from Ref.~\cite{Levai:1997yx}.

We now calculate the operator product expansion (OPE) of the following
correlation function of the vector current $j_\mu=\bar{c} \gamma_\mu c$.
\begin{eqnarray}
\Pi(q^2) & = & \frac{-1}{3q^2} \int d^4x e^{iqx} \langle T [j_\mu(x), j^\mu (0)] \rangle  ,
\end{eqnarray}
Taking the moments
\begin{eqnarray}
M_n(Q_0^2) =\frac{1}{n!} \bigg( -\frac{d}{dQ^2} \bigg)^n \Pi(Q^2)|_{Q^2=Q_0^2},
\end{eqnarray}
we find the following form up to dimension 4 operators \cite{Klingl:1998sr}
\begin{align}
&M_n(\xi)=A^V_n(\xi)\left[ 1+a_n(\xi){\alpha}_s +b_n(\xi){\phi}^4_b +c_n(\xi){\phi}^4_c \right]\\
&{\phi}^4_b=\frac{4\pi^2}{9}\frac{\langle \frac{{\alpha}_s}{\pi}G^2\rangle}{(4{m_c^2})^2}\\
&{\phi}^4_c=\frac{4\pi^2}{3}\frac{G_2}{(4{m_c^2})^2}.
\end{align}

The additions from dimension 6 operators are of the following form
\cite{SuHoungLee}.
\begin{align}
&\Delta M^6_n(\xi)=A^V_n(\xi)\left[ s_n(\xi){\phi}^6_s+x_n(\xi){\phi}^6_x+g_{4n}(\xi){\phi}^6_{g_4} \right]\\
&{\phi}_s^6=\frac{4\pi^2}{3 \cdot 1080} \frac{\langle \frac{\alpha_s}{\pi}G^a_{\mu\nu}G^a_{\mu\nu;\kappa\kappa} \rangle}{(4m_c^2)^3}\\
&{\phi}_x^6=\frac{9 }{2}\frac{4\pi^2}{3 \cdot 1080} \frac{X}{(4m_c^2)^3}\\
&{\phi}_{g_4}^6=10 \frac{4\pi^2}{3 \cdot 1080} \frac{G_4}{(4m_c^2)^3},
\end{align}
where the Wilson coefficients are summarized in Ref.~\cite{SuHoungLee} and we take $\xi=Q_0^2/(4m_c^2)=1$.

Before calculating the mass of $J/\psi$, it is useful to discuss the
effects of adding each contribution from dimension 4 and 6
condensates. The left and right graphs of Fig.~\ref{fig2} show the total
contributions to moments $M_n(\xi)$ from dimension 4 and 6 condensates,
respectively,  divided by the perturbative contribution at $T=0.8, 1.05,
1.2 T_c$; the result at $T=0.8 T_c$ being almost identical to that at
$T=0$.  One notes that at $T=1.05 T_c$, the contributions change signs
in both figures.  This suggests that the finite temperature corrections
become slightly larger than the vacuum condensate value, which shows the
onset of large temperature correction.  The correction at $T=1.2T_c$ are
twice as large as the vacuum value with opposite sign  suggesting  the
need to consider improved method to consider temperature effects.
Furthermore, for each temperature, the moments change sign after $n=21$.
As higher $n$ are more sensitive to higher orders of  power correction,
this signals the onset of the breakdown of the OPE.

Fig.~\ref{fig3} shows the contributions to moments $M_n(\xi)/A^V_n(\xi)$
of each dimension 6 condensates at $T=0.8$(left figure) and
$1.05T_c$(right figure).   As can be seen from the left figure, the
scalar condensate dominates  for any $n$ at $T=0.8T_c$. On the other
hand, the right figure shows that the temperature dependent $\phi_x$ and
$\phi_{g4}$ dominates over the scalar condensate.  Moreover, while the
absolute value of $\phi_{g4}$ is larger than $\phi_x$, they have
opposite signs and tend to cancel the contributions to the moment each
other, as also seen from the right of Fig.~\ref{fig2}.  Again this
signals the onset of large temperature dependence at $T=1.05T_c$.

\begin{figure}[!t]
  \centering
    \includegraphics[width=0.45\textwidth]{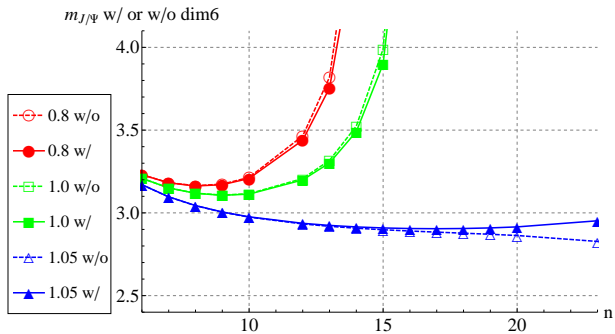}\\
  \caption{$m_{J/\psi}$ at $T=0.8,1.0,1.05T_c$ with or without dimension 6 condensates}
  \label{fig5}
\end{figure}

Let us now look at the mass of $J/\psi$.  Assuming that the imaginary
part of the correlation function is dominated by the lowest pole, the
mass is obtained from
\begin{eqnarray}
m_{J/\psi}^2 = \frac{M_{n-1}}{M_n}-4m_c^2. \label{mjpsi-n}
\end{eqnarray}
Now, we adopt the temperature dependent condensate values as discussed
in the previous section to calculate the temperature dependent mass.

Fig.~\ref{fig4} shows the moment sum rule for the mass as given in Eq.~(\ref{mjpsi-n}).  The left panel of Fig.~\ref{fig4} shows the result when only the  contributions from dimension 4 operators are considered.  The right panel shows the result after contributions from dimension 6 operators are added.
One notes that while the values of the mass shift
do not change much, the stability of the sum rules with dimension 6
condensate improves over the sum rules with dimension 4
condensate. 
Specifically, the instability in the sum rule with dimension 4
condensate at $1.05T_c$ turns into a stable plateau structure by
including dimension 6 condensate, as can be seen in Fig.~\ref{fig5}. 
The stability in $n$ guarantees that the assumptions of the operator product expansion side and phenomenological side are both valid.  Therefore, including the contribution from dimension 6 operators seems to extend the region of stability to slightly higher temperature.  This suggest that the $J/\psi$ will still survive to this temperature.

\section{Summary}

We have introduced a parametrization of the temperature dependence of the
dimension 6  gluon operators based on the temperature dependence of the dimension 4 electric and magnetic condensates extracted from lattice gauge theory.
We then improved the
 previous QCD sum rules for the $J/\psi$ mass near $T_c$ based on
 dimension 4 operators, by including the contribution of the temperature dependent dimension 6  operators to the OPE. We find that the addition extends the stability in the sum rule up to slightly higher temperature of $1.05 T_c$.

\section*{Acknowledgements}

This work was supported by the Korean Research Foundation under Grant Nos. KRF-2011-0020333 and KRF-2011-0030621, and the Grants-in-Aid for Scientific Research on Innovative Areas from MEXT (No. 24105008).

\end{document}